\documentclass{cflowproc}

\usepackage{natbib}

\begin{document}


\shortauthors{KEMNPNER \& DAVID}     
\shorttitle{CHANDRA OBSERVATION OF A576} 

\title{Chandra Observation of the Core of Abell 576}   

\author{Joshua C. Kempner\affilmark{1} 
and Laurence P. David\affilmark{1}}

\affil{1}{Harvard-Smithsonian Center for Astrophysics, 60 Garden St., MS-67, Cambridge, MA 02138}


\begin{abstract}
We present data from a {\it Chandra} observation of the nearby cluster of
galaxies Abell 576.  The core of the cluster shows a significant departure
from dynamical equilibrium.  We show that this core gas is most likely the
remnant of a merging subcluster, which has been stripped of much of its
gas, depositing a stream of gas behind it in the main cluster.  The
unstripped remnant of the subcluster is characterized by a different
temperature, density and metalicity than that of the surrounding main
cluster, suggesting its distinct origin.  Continual dissipation of the
kinetic energy of this minor merger may be sufficient to counteract most
cooling in the main cluster over the lifetime of the merger event.
\end{abstract}


\section{Introduction}
\label{Kempner:intro}

With its low redshift, Abell 576 makes excellent use of the capabilities of
{\it Chandra}, allowing us to examine in detail the very core of the
cluster.  We focus here on the dynamical activity in the core of cluster.
The cluster shows strong evidence, first suggested by \citet{kempner_mgf+96} from
an analysis of the galaxy population, of the remnant core of a small merged
subcluster.  We demonstrate that the X-ray data are consistent with this
picture, and even suggest it as the most likely origin for the
non-equilibrium gas at the center of Abell 576.  In fact, the subcluster
may still be in the process of settling into the center of the main
cluster's potential.

We present 27.9 kiloseconds of {\it Chandra} ACIS-S data (OBSID 3289).
The data have been corrected for CTI and the particle background
reduced using the standard procedure for data taken in Very Faint
mode.  Background corrections were performed using blank sky files
provided in the CALDB.  For the spectroscopic analysis we considered
only data in the range 0.5--8 keV.  The arfs were corrected for the
reduction in quantum efficiency at low energies using the {\it acisabs}
model.

We assume $H_0 = 71$, $\Omega_m = 0.27$, and $\Omega_\Lambda = 0.73$, so
$1\arcsec=0.738$ kpc at the redshift of A576 ($z=0.0377$). All errors are
quoted at 90\% confidence unless otherwise stated.

\section{Brightness Edges}
\label{Kempner:edges}

\begin{figure}
\plotone{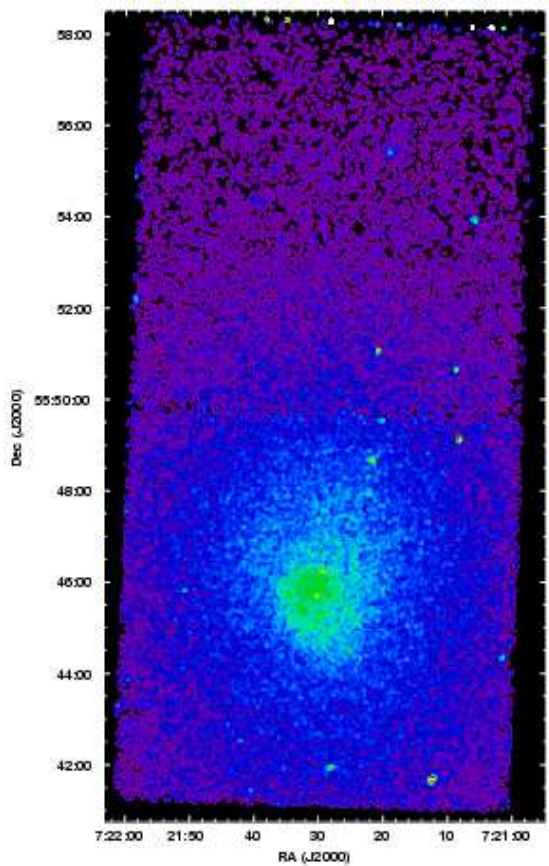}
\figcaption{Gaussian smoothed, exposure-corrected, 0.3--6.0 keV image of
Abell 576.  Both the S2 and S3 chips are shown.  The north and southeast
surface brightness edges are clearly visible; the west edge is less
distinct.
\label{Kempner:img}}
\end{figure}

Figure~\ref{Kempner:img} shows a Gaussian smoothed, exposure-corrected
image of the cluster.  At least two, and perhaps more, surface brightness
edges are visible within the central 50 kpc.  As shown below, they
encompass a region of cool, high-metalicity gas.  This cool gas also
extends in a finger to the north of the cluster core, slightly west of
center.  As we will discuss in the remainder of this section, we believe
this finger of gas to have originated in a small subcluster which is
currently accreting into the center of the main cluster.  The orientation
of the edges are not consistent with gas simply sloshing back and forth in
a more or less fixed potential, which would create parallel edges, but are
more consistent with being the outer edges of a wake of stripped gas left
behind by a merging subcluster.  This hypothesis also neatly explains the
finger of gas to the north, which cannot be easily explained by simple
sloshing.

\begin{figure}
\plotone{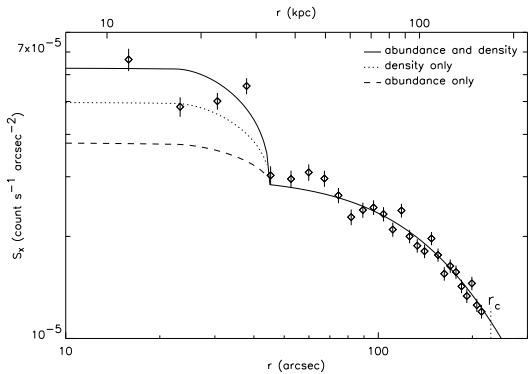}
\figcaption{Surface brightness profile across the north edge.  The solid line
outside 30 kpc is a beta model fit to the data using the parameters derived
by \protect\citet{kempner_mef+95} using the full field of view of the {\it
Einstein} IPC.  Inside 30 kpc, three models for the surface brightness jump
are shown: the solid line indicates the same beta model after adjusting for
the enhancement due to both the density and abundance increases inside the
edge.  The abundance only and density only components are shown as the
dashed and dotted lines, respectively .  For the $\beta$-model, $\beta =
0.64$ and $r_c = 169$ kpc.  The core radius of the model is indicated as
``${\rm r_c}$'' for reference.  The errors bars are 1$\sigma$.
\label{Kempner:north_sbr}}
\end{figure}

\begin{figure}
\plotone{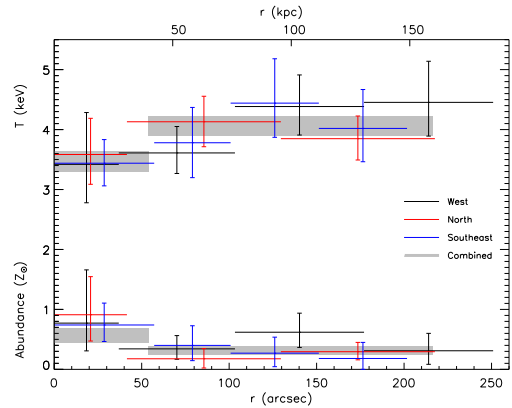}
\figcaption{Spectral profiles in three sectors to the north, southeast, and
west from the cluster center.  The top set of points are temperatures; the
bottom set are abundances.  All error bars are 90\% confidence.
\label{Kempner:spec_profile}}
\end{figure}

We extracted surface brightness and spectral profiles across all three
edges.  The brightness edge 40\arcsec\ north of the peak of the X-ray
emission shows by far the largest jump in surface brightness: a factor of
$1.8\pm0.15$ (1$\sigma$) increase across the discontinuity (see
Figure~\ref{Kempner:north_sbr}).  A large jump in the abundance is also
visible across the discontinuity, while the temperature does not change
significantly (see the red points in Figure~\ref{Kempner:spec_profile}).
At the low temperature of Abell 576, the increased abundance across the
edge has a non-negligible effect on the emissivity of the gas.  This is
illustrated in Figure~\ref{Kempner:north_sbr}, which shows the surface
brightness profile across the north edge.  Outside the brightness edge, the
solid line is a $\beta$-model fit to the data using the core radius and
slope determined from observations with {\it Einstein} \citep{kempner_mgf+96}.
Inside the edge, we use the same $\beta$-model, but we increase the
emissivity by an amount expected from each of three different models: the
dotted line indicates the increased surface brightness due to the increase
in density; the dashed line indicates the increase due to the higher
abundance, and the solid line is the increased brightness due to both
effects.  For all three models we assume spherical symmetry for consistency
with the deprojection analysis.  As the figure demonstrates, neither the
added emissivity from the higher abundance nor that from the increased
density can account entirely for the observed increase in surface
brightness, but the two effects combined reproduce the overall
normalization of the central brightness quite well.

We deprojected the surface brightness, which, when combined with the
temperature and abundance profiles, allowed us to determine deprojected
density and pressure profiles across the edge, under the assumption of
spherical symmetry.  We find a small jump in both the density and the
pressure across the north edge.  The density increases by a factor of
$2.8^{+0.8}_{-1.2}$ while the pressure increases by a factor of $2.4\pm0.5$
(both 1$\sigma$).  In order for the higher density gas to remain confined,
the pressure difference across the edge must be balanced by ram pressure
from motion of the high density gas through the lower density gas.  The
observed pressure difference implies that the higher density gas is moving
through the lower density gas with a velocity of $750 \pm 270$ km s$^{-1}$,
or Mach $0.9 \pm 0.3$ at the sound speed of the lower density gas (both
errors are 1$\sigma$).  The velocity we measure is consistent with
velocities of both merging/accreting subclusters measured in other clusters
\citep[e.g.][]{kempner_mpn+00} and with velocities measured for some ``sloshing''
edges in the cores of relaxed clusters \citep{kempner_mar03}.  On its own, then,
the measured velocity of the north edge is incapable of distinguishing
between these two scenarios for the creation of the non-hydrostatic
features in the cluster core.

To the southeast of the cluster center, a fainter edge is visible in the
image (see Figure~\ref{Kempner:img}).  Another yet fainter edge appears to
the west.  Both of these edges display the same abundance gradient as the
north edge, though in both cases the abundance jump appears to be more of a
gradient than a sharp edge, and is measured with much less significance
(see Figure~\ref{Kempner:spec_profile}).

\section{Suppression of Cooling}
\label{Kempner:cflow}

As Figure~\ref{Kempner:spec_profile} shows, the temperature drops in the
very core of the cluster, that is, inside the brightness edge.  It is
natural to ask, then, if this gas shows any evidence of being multi-phase.
To test this, we fit a spectrum of the gas in the core with a
single-temperature absorbed MEKAL model \citep{kempner_kaa92}, with a MEKAL model
plus a multi-phase MEKAL model (MKCFLOW model), and with the sum of two
MEKAL models.  Application of the F-statistic shows the cooling flow model
to be a marginally better fit than the single-phase model, and the
two-temperature model to be an even better fit, although none is so much
better as to be considered preferred over the others.

\begin{figure*}
\epsscale{1.17}
\plottwo{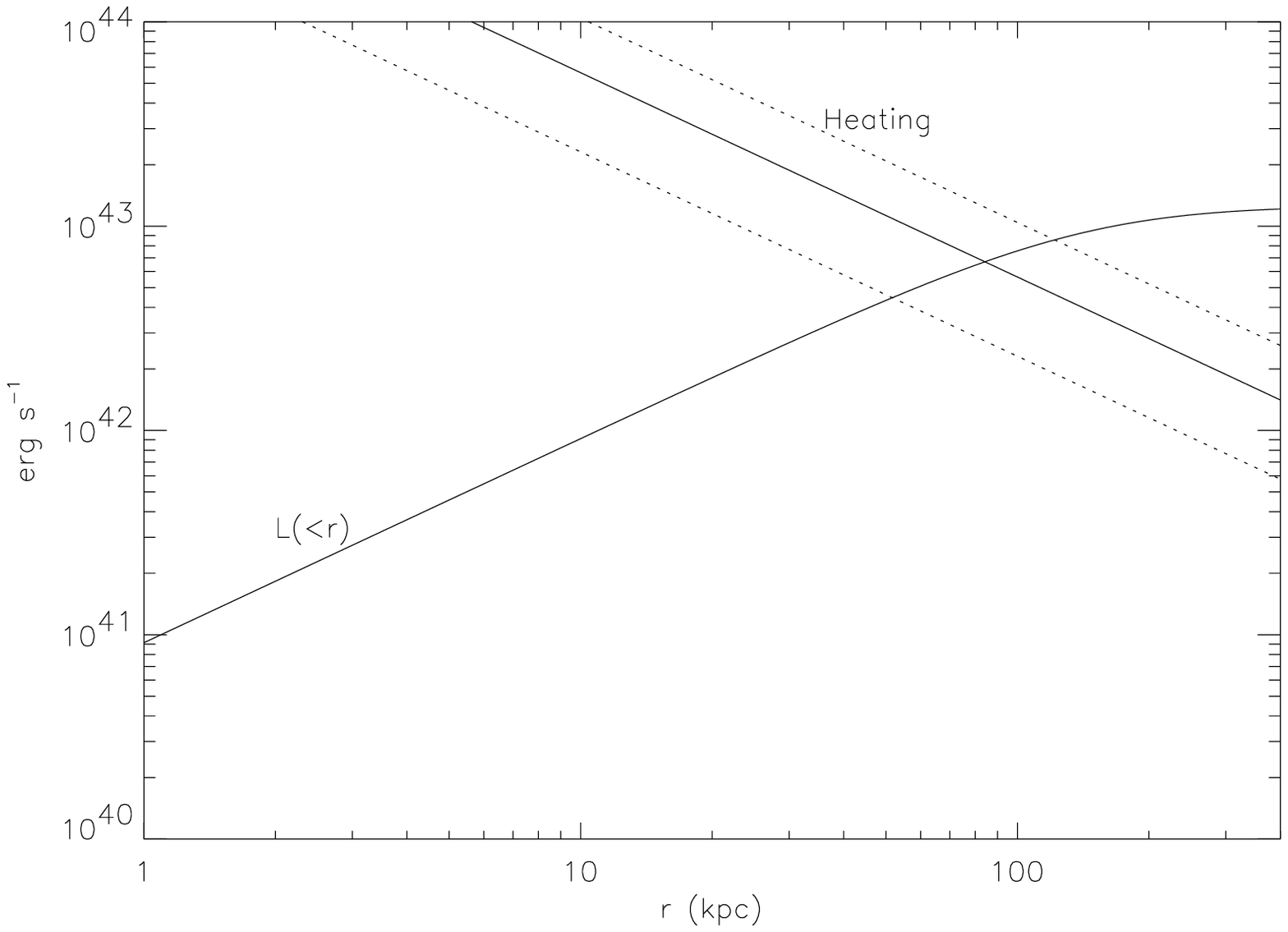}{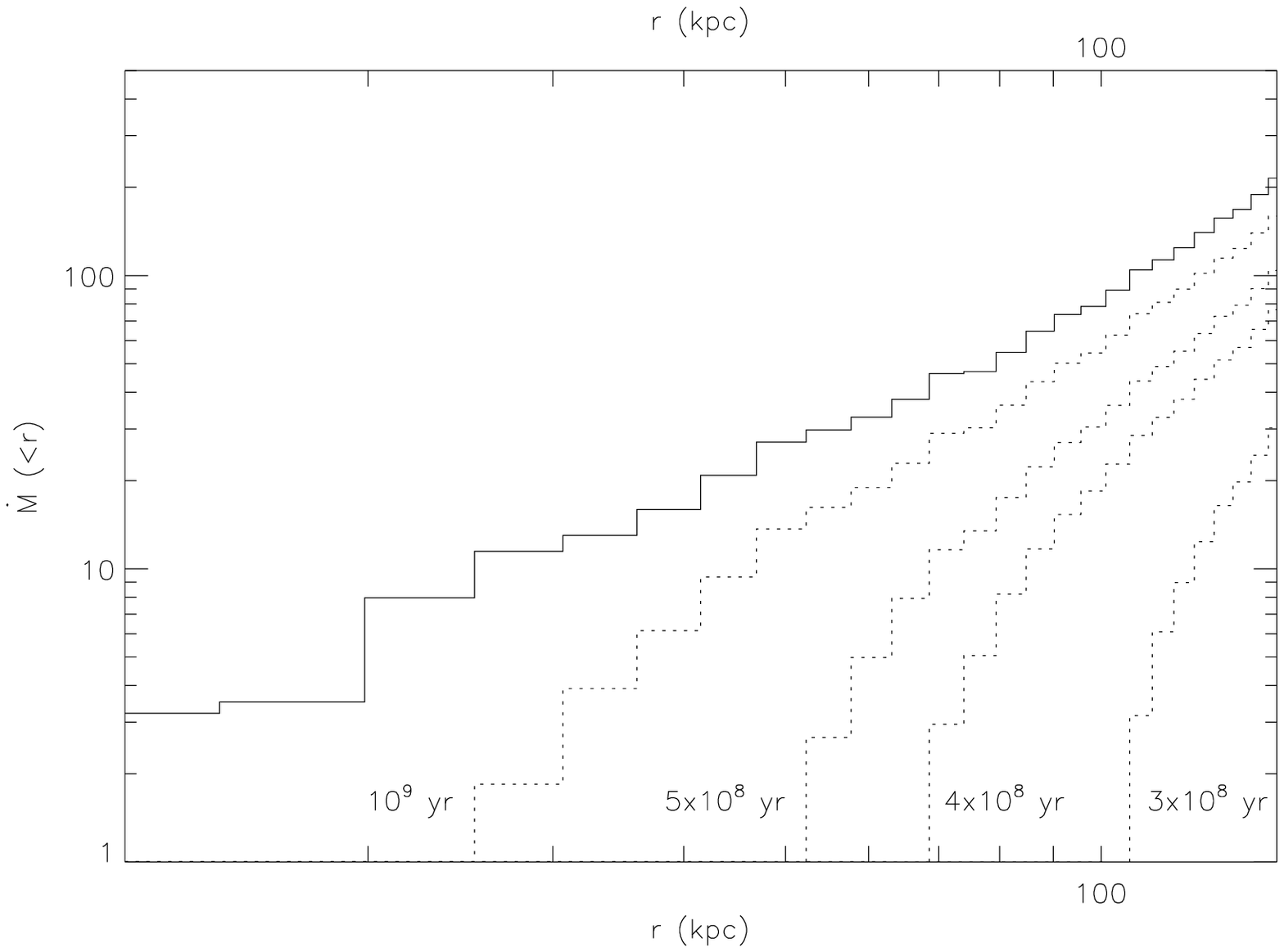}
\figcaption{{\it (a)} Heating rate and luminosity of radiative cooling as a
function of radius.  The exact definition of the heating rate is given in
the text.  The dotted curves indicate the minimum and maximum allowed rates
given the errors on our measurement of the subcluster's velocity.  {\it
(b)} Integrated cooling rate as a function of radius.  The solid line is
the rate in the absence of heating.  The dotted lines are the resulting
integrated cooling rate if the kinetic energy of the cool core is
dissipated over the given timescales.  The dissipated energy is assumed to
heat the gas at a constant rate per volume over the entire region.
\label{Kempner:cooling}}
\end{figure*}

The spectroscopic cooling rate we measure is quite low: an order of
magnitude smaller than the ``classical cooling flow'' accretion rate of
$\dot{M} = M / t_{\rm cool} = 11$ M$_\sun$ yr$^{-1}$ derived from the gas
mass and cooling time in the inner 30 kpc (see
Figure~\ref{Kempner:cooling}).  We should note that the cool component in
the two-temperature model only contributes $\sim$0.1\% to the overall
normalization of the model.  This is consistent with the extremely small
cooling rate measured for the cooling flow model.  In short, we find very
marginal evidence of multi-phase gas in the central cool core.

Given that we observe essentially no cooling in either the core or outside
the core in what is otherwise a quite relaxed cluster, we now explore
whether the dissipation of the kinetic energy of the remnant core is
capable of suppressing cooling at the observed level.  We assume for this
discussion that the merging subcluster explanation of the brightness edges
is correct.

We determined the kinetic energy of the gas inside the north edge using the
velocity of the edge we measured above plus a gas mass determined from the
deprojected density profile inside the edge ($\sim2\times10^{10}~{\rm
M_\sun}$).  We then calculated the rate of energy input from the
dissipation of this kinetic energy over a variety of timescales.
Figure~\ref{Kempner:cooling}a shows this energy dissipation rate compared
to the luminosity due to radiative cooling as a function of radius.  The
timescale used for calculating the heating rate is three crossing times of
the cluster to the given radius at the current velocity of the north edge.
In three crossing times, the moving cool core will have swept up its mass
in gas, reducing its kinetic energy by $3/4$.  We therefore assume
perfectly efficient thermalization of $3/4$ of the kinetic energy over a
timescale equal to three crossing times by the core at its current velocity
in calculating the heating rate.

If we take the point at which the west and southeast edges converge as
indicative of the current orbital radius of the subcluster ($\sim$100 kpc
from the cluster center), the dissipation timescale derived using the above
method is $4 \times 10^8$ years.  As can be seen from
Figure~\ref{Kempner:cooling}b, the heating simply from the dissipation of
the kinetic energy of the subcluster is capable of suppressing cooling by a
factor of more than 4 in the inner 100 kpc over this timescale.  If the
dissipation of the core's kinetic energy is spread out over $10^9$ years,
cooling can still be suppressed at the level observed in the inner 30 kpc.


\acknowledgements
Support for this work was provided by the National Aeronautics and Space
Administration through {\it Chandra} Award Number G01-2131X issued by the {\it
Chandra} X-ray Observatory Center, which is operated by the Smithsonian
Astrophysical Observatory for and on behalf of NASA under contract
NAS8-39073, and by NASA contract NAG5-12933.


\begin{thebibliography}{0}
\expandafter\ifx\csname natexlab\endcsname\relax\def\natexlab#1{#1}\fi
\expandafter\ifx\csname bibnamefont\endcsname\relax
  \def\bibnamefont#1{#1}\fi
\expandafter\ifx\csname bibfnamefont\endcsname\relax
  \def\bibfnamefont#1{#1}\fi
\expandafter\ifx\csname citenamefont\endcsname\relax
  \def\citenamefont#1{#1}\fi
\expandafter\ifx\csname url\endcsname\relax
  \def\url#1{\texttt{#1}}\fi
\expandafter\ifx\csname urlprefix\endcsname\relax\def\urlprefix{URL }\fi
\providecommand{\bibinfo}[2]{#2}
\providecommand{\eprint}[2][]{\url{#2}}

\end{thebibliography}


\begin{thebibliography}{}

\bibitem[{{Kaastra}(1992)}]{kempner_kaa92}
{Kaastra}, J.~S. 1992, {An X-Ray Spectral Code for Optically Thin Plasmas},
  Tech. rep., {Internal SRON-Leiden Report, updated version 2.0}

\bibitem[{{Markevitch} {et~al.}(2000){Markevitch}, {Ponman}, {Nulsen}, {Bautz},
  {Burke}, {David}, {Davis}, {Donnelly}, {Forman}, {Jones}, {Kaastra},
  {Kellogg}, {Kim}, {Kolodziejczak}, {Mazzotta}, {Pagliaro}, {Patel}, {Van
  Speybroeck}, {Vikhlinin}, {Vrtilek}, {Wise}, \& {Zhao}}]{kempner_mpn+00}
{Markevitch}, M., {Ponman}, T.~J., {Nulsen}, P.~E.~J., {Bautz}, M.~W., {Burke},
  D.~J., {David}, L.~P., {Davis}, D., {Donnelly}, R.~H., {Forman}, W.~R.,
  {Jones}, C., {Kaastra}, J., {Kellogg}, E., {Kim}, D.-W., {Kolodziejczak}, J.,
  {Mazzotta}, P., {Pagliaro}, A., {Patel}, S., {Van Speybroeck}, L.,
  {Vikhlinin}, A., {Vrtilek}, J., {Wise}, M., \& {Zhao}, P. 2000, \apj, 541,
  542

\bibitem[{{Markevtich}(2003)}]{kempner_mar03}
{Markevtich}, M. 2003, these proceedings

\bibitem[{{Mohr} {et~al.}(1995){Mohr}, {Evrard}, {Fabricant}, \&
  {Geller}}]{kempner_mef+95}
{Mohr}, J.~J., {Evrard}, A.~E., {Fabricant}, D.~G., \& {Geller}, M.~J. 1995,
  \apj, 447, 8

\bibitem[{{Mohr} {et~al.}(1996){Mohr}, {Geller}, {Fabricant}, {Wegner},
  {Thorstensen}, \& {Richstone}}]{kempner_mgf+96}
{Mohr}, J.~J., {Geller}, M.~J., {Fabricant}, D.~G., {Wegner}, G.,
  {Thorstensen}, J., \& {Richstone}, D.~O. 1996, \apj, 470, 724

\end{thebibliography}

\end{document}